\documentclass[twocolumn,twoside,slac_two]{revtex4}
\usepackage{graphicx}
\usepackage{fancyhdr}
\input babarsym

\newcommand{\dstarstarp}{\ensuremath{D^{**+}}}

\newcommand{\dstarstarz}{\ensuremath{D^{**0}}}
\newcommand{\dstarstar}{\ensuremath{D^{**}}}

\def\Brec  {\ensuremath{\B_{\rm rec'd}}\xspace}
\def\BbX  {\ensuremath{\Bbar_{\rm X \pi}}\xspace}
\def\D        {\ensuremath{D}\xspace}
\def\K        {\ensuremath{K}\xspace}

\def\DsTT{D_{sJ}^*(2317)^+}
\def\DsTO{D_s^{*}(2112)^+}
\def\DsFE{D_{sJ}(2460)^+}
\def\DsTS{D_{s1}(2536)^+}
\def\DsST{D_{s2}(2573)^+}

\def\deltadso{\ensuremath{\Delta m(\Dsop)}\xspace}
\def\dmmean{\ensuremath{\Delta \mu(\Dsop)}\xspace}

\def\Dsop    {\ensuremath{D^{+}_{s1}}\xspace}

\def\Dso     {\ensuremath{D_{s1}}\xspace}
\def\Dst     {\ensuremath{D_{s2}}\xspace}

\newcommand{\kevcc}{\ensuremath{{\mathrm{\,ke\kern -0.1em V\!/}c^2}}\xspace}

\begin{document}

\title{Charm Meson Spectroscopy at \babar\ and CLEO-c}

%

\author{A. Zghiche\\
 From the \babar\ Collaboration}
 \affiliation{ Laboratoire de Physique des Particules,CNRS-IN2P3,
F-74941 Annecy-le-Vieux - France}

\begin{abstract}
  In this mini-review we report on the most recent progress in charm meson spectroscopy. We
discuss the precision measurements performed by the \babar\ and
CLEO-c experiments in the non strange charm meson part and  we
present the newly discovered strange charmed meson excited states.

\end{abstract}

\maketitle

\thispagestyle{fancy}


\section{Introduction}
During the last few years many new $D$, $D_s$, charmonium, and
charmed baryon excited states have been discovered. Some of these
states were not expected theoretically; their masses, widths,
quantum numbers, and decay modes did not fit the existing
spectroscopic classification, which was based mostly on potential
model calculations. The theoretical models had to  be improved and
new approaches have been developed to explain the data; the
possibility of a non-quark-antiquark interpretation of these
states has also been widely discussed. Charmonium, and charmed
baryon excited states results are discussed elsewhere in these
proceedings. In this report an overview of recent results on non
strange charm mesons  production is presented. Then, recent
results on excited $D_{sJ}$ meson production will be presented and
their behavior will be discussed.

\section{Non Strange Charm Mesons}

\subsection{ Measurement of the Absolute Branching Fractions $\boldmath{B
\to D\pi, D^*\pi, D^{**}\pi} $ with a Missing Mass Method}

Our understanding of hadronic \B-meson decays has improved
considerably during the past few years with the development of
models based on the Heavy Quark Effective Theory (HQET), where
collinear~\cite{Bauer2,Bauer3} or $k_T$~\cite{Botts,Li}
factorization theorems are considered. Models such as the
QCD-improved Factorization (QCDF)~\cite{Beneke,Neubert1} and the
Soft Collinear Effective Theory (SCET)~\cite{Bauer2,Bauer1} use
the collinear factorization, while the perturbative QCD (pQCD)
 approach~\cite{Keum,Kurimoto}
uses the $k_T$ factorization. In these models the amplitude of the
$B\to D^{(*)}\pi$ two-body decay carries information about the
 difference $\delta$ between the
strong-interaction phases of the two isospin amplitudes $A_{1/2}$
and $A_{3/2}$ that contribute~\cite{Rosner,Chiang}. A non-zero
value of $\delta$ provides a measure of the departure from the
heavy-quark limit and the importance of the final-state
interactions in the $D^{(*)}\pi$ system. With the measurements by
the \babar\ \cite{ref:babard0pi0} and
  BELLE \cite{ref:belled0pi0} experiments  of the color-suppressed  $B$ decay
$\Bzb\to\D^{(*)0}\piz$ providing evidence for a sizeable value of
$\delta$, an improved measurement of the color-favored decay
amplitudes ($\Bm\to\D^{(*)0}\pim$  and $\Bzb\to\D^{(*)+}\pim$) is
of renewed interest. In addition, the study of
 $B$~decays into  $D$, $D^{*}$, and $D^{**}$  mesons will
allow tests of  the spin symmetry~\cite{Mannel,NRSX,Mantry,Jugeau}
imbedded in HQET and  of  non-factorizable corrections~\cite{Blok}
that have been assumed to be negligible in the case of the excited
states $D^{**}$ \cite{Neubert2}.

A measurement of the branching fractions is presented for the
decays
 \Bm\to\Dz\pim, \Dstarz\pim, \dstarstarz\pim and \Bzb\to\Dp\pim,
 \Dstarp\pim, \dstarstarp\pim~\cite{footnote}
with a missing mass method, based on a sample of 231 million
\FourS\to\BB pairs collected by the \babar\ detector at the PEP-II
\epem collider. One of the B mesons is fully reconstructed and the
other one decays to a reconstructed $\pi$ and a companion charmed
meson identified by its recoil mass, inferred by the kinematics of
the two body \B decay. This method, compared to the previous
exclusive measurements~\cite{cleo1}, does not imply that the
\FourS decays into \Bp and \Bz with equal rates, nor rely on the
$D$, $D^*$, or $D^{**}$ decay branching fractions. The number of
fully reconstructed B mesons \Brec is extracted from a fit to its
mass distribution. In the decay $\FourS \to \Brec \BbX$ where
$\BbX $ is the recoiling $\Bbar$ which decays into $\pim X$, the
invariant mass of the $X$ system is derived from the missing
4-momentum $p_X$ applying the energy-momentum conservation:
\begin{eqnarray}
p_X =p_{\FourS}-p_{\Brec}-p_{\pim}. \nonumber
\end{eqnarray}
The 4-momentum of the \FourS, $p_{\FourS}$, is computed from the
beam energies and $p_\pi$ and $p_{\Brec}$ are the measured
4-momenta  of the pion and of the reconstructed $\Brec$,
respectively. The $\Brec$ energy is constrained by the beam
energies.
 The  $\Bbar\to D \pim$, $\Bbar\to D^* \pim$, or
$\Bbar\to D^{**} \pim$ signal yields peak at the $D$, $D^*$, and
$D^{**}$ masses in the missing mass spectrum, respectively. The
signal yield of the different modes, is extracted from the missing
mass spectra. The $\D\pi$ and $\Dstar\pi$ signal yields are
extracted  by a $\chi^2$ fit to the background subtracted missing
mass distribution in the range $1.65-2.20~\gevcc$. The $D^{**}$
yield is obtained by counting the candidates in excess in the
missing mass range $2.2-2.8~\gevcc$. This range is chosen in order
to keep most of the excess and no further assumption on
$\dstarstar$ resonance composition is made. The following
branching fractions~\cite{dss} are measured:
 \begin{eqnarray}
\BR(\Bm \to \Dz\pi^-)&  =&  (4.49\pm 0.21 \pm 0.23)\times 10^{-3} \nonumber  \\
\BR(\Bm \to \Dstarz\pi^-) & =& (5.13 \pm 0.22 \pm 0.28)\times 10^{-3}  \nonumber  \\
\BR(\Bm \to \dstarstarz\pi^-) & =& (5.50 \pm 0.52 \pm 1.04)\times 10^{-3}\nonumber \\
 \BR(\Bzb \to \Dp\pi^-)& =&  (3.03 \pm 0.23 \pm 0.23)\times 10^{-3}\nonumber  \\
 \BR(\Bzb \to \Dstarp\pi^-)& =& (2.99 \pm 0.23 \pm 0.24)\times 10^{-3} \nonumber  \\
  \BR(\Bzb \to \dstarstarp\pi^-) &= & (2.34 \pm 0.65 \pm 0.88)\times 10^{-3} \nonumber
\end{eqnarray}
and the branching ratios:
\begin{eqnarray}
\BR(\Bm\to\Dstarz\pi^-)/\BR(\Bm\to\Dz\pi^-)=1.14 \pm 0.07\pm0.04\  \nonumber\\
\BR(\Bm\to\dstarstarz\pi^-)/\BR(\Bm\to\Dz\pi^-)=1.22\pm0.13\pm0.23\ \nonumber\\
\BR(\Bzb\to\Dstarp\pi^-)/\BR(\Bzb\to\Dp\pi^-)=0.99\pm 0.11\pm0.08\ \nonumber\\
\BR(\Bzb\to\dstarstarp\pi^-)/\BR(\Bzb\to\Dp\pi^-)=0.77\pm0.22\pm0.29\
\nonumber
\end{eqnarray}
The first uncertainty is statistical and the second is systematic.
This result is published~\cite{zgh}.

\subsection{Precision Measurement of \Dz mass by CLEO-c}
The \Dz (c\ubar) and \Dpm(d\cbar,c\dbar) mesons form the ground
states of the open charm system. The knowledge of their masses is
important for its own sake, but a precision determination of the
\Dz mass has become more important because of the recent discovery
of a narrow state known as
X(3872)~\cite{choi,acosta,abazov,aubert}. Many different
theoretical models have been
proposed~\cite{eichten,close,seth,swanson} to explain the nature
of this state, whose present average of measured masses is
$M(X)=3871.2 \pm 0.5$ \mev~\cite{pdg}. A provocative and
challenging theoretical suggestion is that X(3872) is a loosely
bound molecule of \Dz and \Dstarz mesons~\cite{swanson}. This
suggestion arises mainly from the closeness of  $M[X(3872)]$ to
$M(\Dz)+M(\Dstarz)=2M(\Dz)+[M(\Dstarz)-M(\Dz)]=2(1864.1 \pm 1.0) +
(142.12 \pm 0.07) \mev = 3870.32\pm 2.0$ \mev based on the
PDG~\cite{pdg} average value of the measured \Dz mass,
$M(\Dz)=1864.1 \pm 1.0$ \mev. This gives the binding energy of the
proposed molecule, $E_b$[X(3872)]= M(\Dz)+M(\Dstarz)- M[X(3872)]=
$-0.9\pm 2.1$ \mev. Although the negative value of the binding
energy would indicate that X(3872) is not a bound state of \Dz and
\Dstarz, its $\pm 2.1$ \mev error does not preclude this
possibility. It is necessary to measure the masses of both \Dz and
X(3872) with much improved precision to reach a firm conclusion.
Recently, CLEO-c reported  a precision measurement of the \Dz
mass, and provided a more constrained value of the binding energy
of X(3872) as a molecule. Several earlier measurements of the \Dz
mass exist. The PDG~\cite{pdg} resulting average \Dz mass is based
on the measured \Dz masses as M(\Dz)$_{AVG}=1864.1\pm 1.0$ \mev.
They also list a fitted mass, M(\Dz)$_{FIT}=1864.5 \pm  0.4$ \mev,
based on the updated results of measurements of \Dpm , \Dz,
\ensuremath\D$^{\pm}_s$, \ensuremath\D$^{*\pm}$, \Dstarz, and
\ensuremath\D$^{*\pm}_s$ masses and mass differences. In its
recent measurement, CLEO-c analyzes 281 pb$^{-1}$ of \epem
annihilation data taken at the $\Psi$(3770) resonance at the
Cornell Electron Storage Ring (CESR) with the CLEO-c detector to
measure the \Dz mass using the reaction $\Psi(3770) \to \Dz\Dzb$,
with $\Dz \to \KS\Phi$, $\KS \to \pipi$ and $\Phi \to \KpKm$. The
choice of the $\Dz \to \KS\Phi$ mode is motivated by the
determination of the \Dz mass not depending on the precision of
the determination of the beam energy. Since M($\Phi$)+M(\KS)=1517
\mev is a substantial fraction of M(\Dz), the final state
particles have small momenta and the uncertainty in their
measurement makes a small contribution to the total uncertainty in
M(\Dz). This consideration favors $\Dz \to \KS\Phi$ over the more
prolific decays $\Dz \to \Km\pip$ and $\Dz \to \Km\pip\pim\pip$ in
which the decay particles have considerably larger momenta and
therefore greater sensitivity to the measurement uncertainties. An
additional advantage of the $\Dz \to \KS\Phi$ reaction is that in
fitting for M(\Dz) the mass of \KS can be constrained to its value
which is known with precision~\cite{pdg}. The final result of this
measurement is M(\Dz) = 1864.847 $\pm$ 0.150 (stat) $\pm$ 0.095
(syst) \mev.
 Adding the errors in quadrature,  M(\Dz)=1864.847 $\pm$ 0.178
\mev is obtained. This is significantly more precise than the
current PDG average~\cite{pdg}. This result for M(\Dz) leads to
M(\Dz\Dstarz)= 3871.81 $\pm$ 0.36 \mev. Thus, the binding energy
of   X(3872)  as a \Dz\Dstarzb molecule is $E_b$ = (3871.81 $\pm$
0.36) - (3871.2 $\pm$ 0.5) = +0.6 $\pm$ 0.6 \mev. This result
provides a strong constraint for the theoretical predictions for
the decays of X(3872)  if it is a \Dz\Dstarzb
molecule~\cite{swanson}. The error in the binding energy is now
dominated by the error in the X(3872)  mass measurement, which
will hopefully improve as the results from the analysis of larger
luminosity data from various experiments become available. This
analysis is published~\cite{cleoc}.

\section{Strange charm mesons}

\begin{figure}
\includegraphics[width=\linewidth]{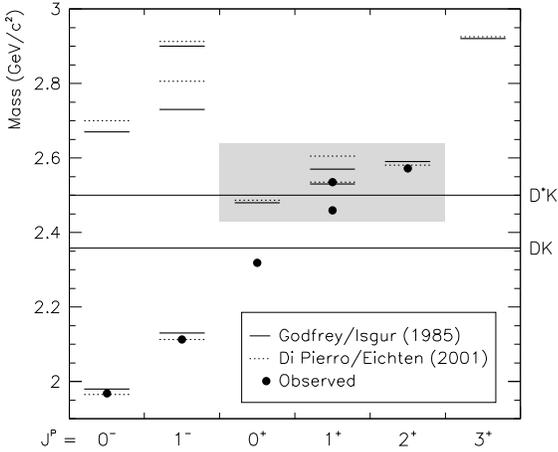}
\vskip -0.25in \caption{\label{fg:godfreyisgur}The $c\bar{s}$
meson spectrum, as predicted by Godfrey and Isgur~\cite{Godfrey1}
(solid lines) and Di Pierro and Eichten~\cite{DiPierro} (dashed
lines) and as observed by experiment (points). The $DK$ and $D^*K$
mass thresholds are indicated by the horizontal lines spanning the
width of the plot. }
\end{figure}

Much of the theoretical work on the $c\bar{s}$ system has been
performed in the limit of heavy $c$ quark mass using potential
models~\cite{Godfrey1,Godfrey2,Isgur,DiPierro} that treat the
$c\bar{s}$ system much like a hydrogen atom. Prior to the
discovery of the $\DsTT$ meson, such models were successful at
explaining the masses of all known $D$ and $D_s$ states and even
predicting, to good accuracy, the masses of many $D$ mesons
(including the $\DsTS$ and $\DsST$) before they were observed (see
Fig.~\ref{fg:godfreyisgur}). Several of the predicted $D_s$ states
were not confirmed experimentally, notably the lowest mass
$J^P=0^+$ state (at around 2.48~\gevcc) and the second lowest mass
$J^P=1^+$ state (at around 2.58~\gevcc). Since the predicted
widths of these two states were large, they would be hard to
observe, and thus the lack of experimental evidence was not a
concern.

\subsection{$\DsTT$ and $\DsFE$}

The $\DsTT$ meson has been observed in the decay
$\DsTT\to\Ds\piz$~\cite{Aubert1,Besson,Abe,Krokovny,Aubert2}. The
mass is measured to be around $2.32$~\gevcc, which is below the
$DK$ threshold. Thus, this particle is forced to decay either
electromagnetically, of which there is no experimental evidence,
or through the observed isospin-violating $\Ds\piz$ strong decay.
The intrinsic width is small enough that only upper limits have
been measured (the best limit previous to this analysis being
$\Gamma<4.6$~\mev\  at 95\% CL as established by
BELLE~\cite{Abe}). If the $\DsTT$ is the missing $0^+$ $c\bar{s}$
meson state, the narrow width could be explained by the lack of an
isospin-conserving strong decay channel. The low mass (160~\mevcc\
below expectations) is more surprising and has led to the
speculation that the $\DsTT$ does not belong to the $\Ds$ meson
family at all but is instead some type of exotic particle, such as
a four-quark state~\cite{Barnes}.

The $\DsFE$ meson has been observed decaying to
$\Ds\piz\gamma$~\cite{Besson,Aubert1,Abe,Krokovny,Aubert2},
$\Ds\pip\pim$~\cite{Abe}, and
$\Ds\gamma$~\cite{Abe,Krokovny,Aubert2}. The intrinsic width is
small enough that only upper limits have been measured (the best
limit previous to this analysis being $\Gamma<5.5$~\mev\  at 95\%
CL as established by BELLE~\cite{Abe}). The $\Ds\gamma$ decay
implies a spin of at least one, and so it is natural to assume
that the $\DsFE$ is the missing $1^+$ $c\bar{s}$ meson state. Like
the $\DsTT$, the $\DsFE$ is substantially lower in mass than
predicted for the normal $c\bar{s}$ meson. This suggests that a
similar mechanism is deflating the masses of both mesons, or that
both the states belong to the same family of exotic particles.

The spin-parity of the $\DsTT$ and $\DsFE$ mesons has not been
firmly established. The decay mode of the $\DsTT$ alone implies a
spin-parity assignment from the natural $J^P$ series
$\{0^+,1^-,2^+,\dots\}$, assuming parity conservation. Because of
the low mass, the assignment $J^P=0^+$ seems most reasonable,
although experimental data have not ruled out higher spin. It is
not clear whether electromagnetic decays such as $\DsTO\gamma$ can
compete with the strong decay to $\Ds\piz$, even with isospin
violation. Thus, the absence of experimental evidence for
radiative decays such as $\DsTT\to\DsTO\gamma$ is not conclusive.

Experimental evidence for the spin-parity of the $\DsFE$ meson is
somewhat stronger. The observation of the decay to $\Ds\gamma$
alone rules out $J=0$. Decay distribution studies in $B\to\DsFE
D_s^{(*)-}$~\cite{Krokovny,Aubert2} favor the assignment $J=1$.
Decays to either $\Ds\piz$, $D^0 K^+$, or $D^+K^0$ would be
favored if they were allowed. Since these decay channels are not
observed, this suggests, when combined with the other
observations, the assignment $J^P = 1^+$. In this case, the decay
to $\DsTT\gamma$ is allowed, but it may be small in comparison to
the $\Ds\gamma$ decay mode.

An updated analysis of the $\DsTT$ and $\DsFE$ mesons using
232~${\rm fb}^{-1}$ of $e^+e^-\to c\bar{c}$ data is presented
here. Established signals from the decay $\DsTT\to\Ds\piz$ and
$\DsFE\to\Ds\piz\gamma$, $\Ds\gamma$, and $\Ds\pip\pim$ are
confirmed. A detailed analysis of invariant mass distributions of
these final states including consideration of the background
introduced by reflections of other $c\bar{s}$ decays produces the
following mass values:
\begin{eqnarray}
m(\DsTT) = (2319.6 \pm   0.2  \pm   1.4)\mevcc \nonumber\\
m(\DsFE) =(2460.1 \pm  0.2  \pm   0.8) \mevcc,\nonumber
\end{eqnarray}
where the first error is statistical and the second systematic.
Upper 95\% CL limits of $\Gamma < 3.8$~\mev\  and $\Gamma <
3.5$~\mev\ are calculated for the intrinsic $\DsTT$ and $\DsFE$
widths. All results are consistent with previous measurements.

The following final states are investigated: $\Ds\piz$,
$\Ds\gamma$, $\DsTO\piz$, $\DsTT\gamma$, $\Ds\piz\piz$,
$\DsTO\gamma$, $\Ds\gamma\gamma$, $\Ds\pipm$, and $\Ds\pip\pim$.
No statistically significant evidence of new decay modes is
observed. The following branching ratios are measured:

\begin{eqnarray}
 \frac{\mathcal B(\DsFE\to\Ds\gamma)}{\mathcal
B(\DsFE\to\Ds\piz\gamma)}
= 0.337 \pm   0.036  \pm   0.038 \nonumber\\
\frac{\mathcal B(\DsFE\to\Ds\pip\pim)}{\mathcal
B(\DsFE\to\Ds\piz\gamma)} = 0.077 \pm   0.013  \pm    0.008
,\nonumber
\end{eqnarray}
where the first error is statistical and the second systematic.
The data are consistent with the decay $\DsFE\to\Ds\piz\gamma$
proceeding entirely through $\DsTO\piz$.

Since the results presented here are consistent with $J^P=0^+$ and
$J^P = 1^+$ spin-parity assignments for the $\DsTT$ and $\DsFE$
mesons, these two states remain viable candidates for the lowest
lying $p$-wave $c\bar{s}$ mesons. The lack of evidence for some
radiative decays, in particular $\DsTT\to\DsTO\gamma$ and
$\DsFE\to\DsTO\gamma$, are in contradiction with this hypothesis
according to some calculations, but large theoretical
uncertainties remain. No state near the $\DsTT$ mass is observed
decaying to $\Ds\pipm$. If charged or neutral partners to the
$\DsTT$ exist (as would be expected if the $\DsTT$ is a four-quark
state), some mechanism is required to suppress their production in
$e^+e^-$ collisions. This analysis is realized in inclusive
$c\bar{c}$ production using 232~${\rm fb}^{-1}$ of data collected
by the \babar\   experiment near $\sqrt{s} = 10.6$~\gev and  is
published in ~\cite{ds0}.


\subsection{The $\Dso(2536)^{+}$ Case}

For a complete understanding of the charmed strange meson
spectrum, a comprehensive knowledge of the parameters of all known
\Ds mesons is mandatory. In this part of the presentation, a
precision measurement of the mass and the decay width of the meson
 $\Dso(2536)^{+}$ is presented. The mass is currently reported by the PDG with a
 precision of $0.6 \mevcc$, while only an upper limit of $2.3 \mevcc$ is given for the
 decay width \cite{pdg}. These values are based on measurements with 20 times
 fewer reconstructed \Dsop candidates compared to this one. The \babar\ experiment,
 in addition to its excellent tracking and vertexing capabilities, provides a rich source
 of charmed hadrons, enabling an analysis of the \Dsop with high statistics and small errors.

Since the uncertainty of the \Dstarp mass is large ($0.4 \mevcc$
\cite{pdg}), a measurement of the mass difference defined by
\begin{eqnarray}
\deltadso = m(\Dsop) - m(\Dstarp) - m(\KS), \nonumber
\end{eqnarray}
is performed. Additionally, due to the correlation between the
masses, the \Dsop signal in the mass difference spectrum is much
more narrow than the one from the \Dsop mass spectrum alone
leading to a high precision measurement of the mass and the decay
width of the meson $\Dso(2536)^{+}$ using the decay mode $\Dsop
\rightarrow \Dstarp\KS$. The mass difference between \Dsop and
$\Dstarp\KS$ for the two reconstructed decay modes is measured to
be
\begin{center}
$\dmmean_{K4\pi} = 27.209 \pm 0.028 \pm 0.031 \mevcc$,\\
$\dmmean_{K6\pi} = 27.180 \pm 0.023 \pm 0.043 \mevcc$,\\
\end{center}
with the first error denoting the statistical uncertainty and the
second one the systematic uncertainty. These results correspond to
a relative error of $0.15\%$ for the mass difference. This lies
within the range of precision achievable with the \babar\
detector: the \jpsi mass has been reconstructed with a relative
error of $0.05\%$~\cite{babar}.

Combining the results, while taking the systematic errors
including the uncertainties of the \Dstarp mass ($\pm 0.4\mevcc$)
and of the \KS mass ($\pm 0.022 \mevcc$) into account, yields
 a final value for the \Dsop mass of
\begin{center}
$m(\Dsop) = 2534.85 \pm 0.02 \pm 0.40 \mevcc$,\\
\end{center}
while the PDG value for the mass is given as $2535.35 \pm 0.34 \pm
0.50 \mevcc$. The error on the measured \Dsop mass is dominated by
the uncertainty of the \Dstarp mass. The mass difference between
the \Dsop and the \Dstarp follows from these results as
\begin{center}
$\Delta m = m(\Dsop) - m(\Dstarp) = 524.85 \pm 0.02 \pm 0.04 \mevcc$.\\
\end{center}
The decay width is measured to be
\begin{center}
$\Gamma(\Dsop)_{K4\pi} = 1.112 \pm 0.068 \pm 0.131 \mevcc$,\\
$\Gamma(\Dsop)_{K6\pi} = 0.990 \pm 0.059 \pm 0.119 \mevcc$.\\
\end{center}
The final combined value for decay width is
\begin{center}
$\Gamma(\Dsop) = 1.03 \pm 0.05 \pm 0.12 \mevcc$.\\
\end{center}
The result for the mass difference $\Delta m = m(\Dsop) -
m(\Dstarp)$ represents an
 improvement in precision by a factor of 14 compared with the current PDG value
  of $525.3 \pm 0.6 \pm 0.1 \mevcc$. It deviates by $1\sigma$ from the
  larger PDG value. The precision achieved is comparable with other recent high precision
  analyses performed at \babar\, like the $\Lambda_{c}$ mass measurement
  (\mbox{$m(\Lambda_{c}) = 2286.46 \pm 0.04 \pm 0.14 \mevcc$)~\cite{bap}}.
Furthermore, this analysis presents for the first time a direct
measurement of the \Dsop decay width with small errors rather than
just an upper limit, which is currently stated by the PDG as $2.3
\mevcc$. This analysis is also realized in inclusive $c\bar{c}$
production using 232~${\rm fb}^{-1}$ of data collected by the
\babar\   experiment near $\sqrt{s} = 10.6$~\gev and is detailed
in~\cite{ds1p}.

\subsection{$\Dst(2573)^{+}$ and New Strange Charmed Mesons }

Here, a new $c\overline{s}$ state and a broad structure observed
in the decay channels $D^0\Kp$ and $D^+K^0_S$ are reported. This
analysis is based on a 240~${\rm fb}^{-1}$ inclusive $c\bar{c}$
data sample recorded near the $\Upsilon(4S)$ resonance by the
\babar\ detector at the \pep2 asymmetric-energy $e^+e^-$ storage
rings.

Three inclusive processes~\cite{footnote} are reconstructed:
\begin{eqnarray}
e^+ e^- &\to& D^0 K^+ X, D^0 \to K^- \pi^+\\
e^+ e^- &\to& D^0 K^+ X, D^0 \to K^- \pi^+ \pi^0\\
e^+ e^- &\to& D^+ K^0_S X, D^+ \to K^- \pi^+ \pi^+, K^0_S \to
\pi^+ \pi^-
\end{eqnarray}
\begin{figure*}
\vskip -0.15in
\includegraphics[width=1.0\linewidth]{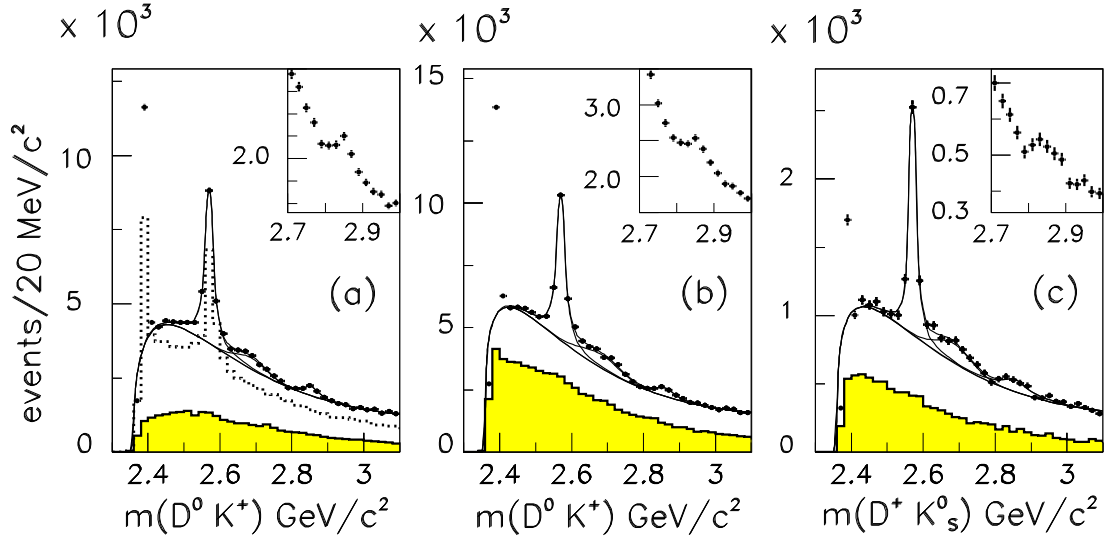}
\vskip -0.15in \caption{\label{fig:fig2} The $D K$ invariant mass
distributions for (a) $D^0_{K^-\pi^+} K^+$, (b)
$D^0_{K^-\pi^+\pi^0} K^+$ and (c) $D^+_{K^-\pi^+\pi^+} K^0_s$. The
shaded histograms are for the $D$-mass sideband regions. The
dotted histogram in (a) is from $e^+ e^- \to c \bar c$ Monte Carlo
simulations incorporating previously known $D_s$ states with an
arbitrary normalization. The insets show an expanded view of the
2.86 \gevcc region. The solid curves are the fitted background
threshold functions from the three separate fits. }
\end{figure*}

Selecting events in the $D$ signal regions, Fig.~\ref{fig:fig2}
shows the $D^0 K^+$ invariant mass distributions for channels (1)
and (2), and the $D^+ K^0_S$ invariant mass distribution for
channel (3).  To improve mass resolution, the nominal $D$ mass and
the reconstructed 3-momentum
are used to calculate the $D$ energy for channels (1) and (3).
Since channel (2) has a poorer $D^0$ resolution, each $\Km \pi^+
\pi^0$ candidate is kinematically fit with a $D^0$ mass constraint
and a $\chi^2$ probability greater than 0.1\% is required.

The fraction of events having more than one $DK$ combination per
event is 0.9\% for channels (1) and (3) and 3.4\% for channel (2).
In the following, the term reflection will be used to describe
enhancements produced by two or three body decays of narrow
resonances where one of the decay products is missed.

The three mass spectra in Fig.~\ref{fig:fig2} present similar
features:
\begin{itemize}
\item{} A single bin peak at 2.4 \gevcc due to a reflection from
the decays of the $D_{s1}(2536)^+$ to $D^{*0} K^+$ or $D^{*+}
K^0_S$ in which the $\pi^0$ or $\gamma$ from the $D^*$ decay is
missed. This state, if $J^P=1^+$, cannot decay to $D K$. \item{} A
prominent narrow signal due to the $D_{s2}(2573)^+$. \item{} A
broad structure peaking at a mass of approximately 2.7~\gevcc.
\item{} An enhancement around 2.86~\gevcc. This can be seen better
in the expanded views shown in the insets of Fig.~\ref{fig:fig2}.
\end{itemize}

Different background sources are examined: combinatorial, possible
reflections from $D^*$ decays, and particle misidentification.

Backgrounds come both from events in which the candidate $D$ meson
is correctly identified  and from events in which it is not. The
first case can be studied combining a reconstructed  $D$ meson
with a kaon from another $\bar D$ meson in the same event, using
data with fully reconstructed $D \bar D$ pairs or Monte Carlo
simulations. No signal near 2.7 or 2.86~\gevcc is seen in the $DK$
mass plots for these events. The second case can be studied using
the $D$ mass sidebands.  The shaded regions in Fig.~\ref{fig:fig2}
show the $DK$ mass spectra for events in the $D$ sideband regions
normalized to the estimated background in the signal region. No
prominent structure is visible in the sideband mass spectra. The
dotted histogram in (a) is from $e^+ e^- \to c \bar c$ Monte Carlo
simulations incorporating previously known $D_s$ states with an
arbitrary normalization.

The possibility that the features at 2.7 and 2.86~\gevcc could be
a reflection from $D^*$ or other higher mass resonances is
considered. Candidate $DK$ pairs where the $D$ is a $D^*$-decay
product are identified by forming $D\pi$ and $D\gamma$
combinations and requiring the invariant-mass difference between
one of those combinations and the $D$ to be within $\pm 2\sigma$
of the known $D^*-D$ mass difference. No signal near 2.7 or
2.86~\gevcc is seen in the $DK$ mass plots for these events.
Events belonging to these possible reflections (except for the
$D^{*0} \to D^0 \gamma$ events, which could not be isolated
cleanly) have been removed from the mass distributions shown in
Fig.~\ref{fig:fig2} (corresponding to $\approx$8\% of the final
sample).

\begin{figure*}
\begin{center}
\vskip -0.15in
\includegraphics[width=1.0\linewidth]{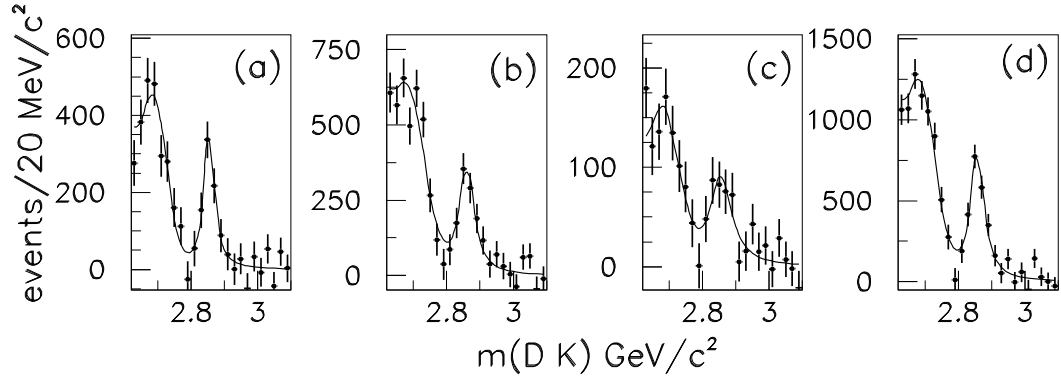}
\end{center}
\vskip -0.15in \caption{\label{fig:fig3} Fitted
background-subtracted $D K$ invariant mass distributions for (a)
$D^0_{K^-\pi^+} K^+$, (b) $D^0_{K^-\pi^+\pi^0} K^+$, (c)
$D^+_{K^-\pi^+\pi^+}K^0_s$, and (d) the sum of all modes in the
2.86 \gevcc mass region. The curves are the fitted functions
described in the text. }
\end{figure*}

The presence of resonant structures can be visually enhanced by
subtracting the fitted background threshold function from the
data. Fig.~\ref{fig:fig3} shows the background-subtracted
$D^0_{K^-\pi^+} K^+$, $D^0_{K^-\pi^+\pi^0} K^+$, and
$D^+_{K^-\pi^+\pi^+} K^0_s$ invariant mass distributions in the
2.86 \gevcc mass region. Fig.~\ref{fig:fig3}(d) shows the sum of
the three mass spectra.

In the following,  the structure in the 2.86~\gevcc mass region is
labelled $D_{sJ}(2860)^+$  and the one  in the 2.7~\gevcc mass
region is labelled  $X(2690)^+$. The three $D K$ mass spectra
shown in Fig.~\ref{fig:fig2} from 2.42 \gevcc to 3.1 \gevcc
(excluding the $D_{s1}(2536)^+$ reflection) are first fitted
separately using a binned $\chi^2$ minimization. The background
for the three $DK$ mass distributions is described by a threshold
function: $(m - m_{\rm th})^{\alpha}~e^{-\beta m-\gamma m^2-\delta
m^3}$ where $m_{\rm th}=m_{D}+m_K$. A fit to the Monte Carlo
distribution shown in Fig.~\ref{fig:fig2}(a) using this background
expression and one spin-2 relativistic Breit-Wigner for the
$D_{s2}(2573)^+$ gives a good  32 \% $\chi^2$ probability. In the
fit to the data, the $D_{s2}(2573)^+$ and $D_{sJ}(2860)^+$ peaks
are described with relativistic Breit-Wigner lineshapes where
spin-2 is assumed for the $D_{s2}(2573)^+$ and spin 0 is used for
the $D_{sJ}(2860)^+$. The $D_{sJ}(2860)^+$ parameters are found
insensitive to the choice of the spin. The best description of the
$X(2690)^+$ structure is obtained using a Gaussian distribution.
The fits give consistent values for the parameters of the three
structures.

When the three mass distributions of Fig.~\ref{fig:fig2} are
fitted simultaneously, the resulting resonance parameters are also
found consistent with those obtained with previous separate fits.
 For $D_{s2}(2573)^+$ resonance, mass
and width are:
\begin{eqnarray*}
m(D_{s2}(2573)^+) = (2572.2 \pm 0.3 \pm 1.0) \ \mevcc \\
\Gamma(D_{s2}(2573)^+)=(27.1 \pm 0.6 \pm 5.6) \ \mevcc,
\end{eqnarray*}
where the first errors are statistical and the second systematic.
For the new states, the following values are obtained:
\begin{eqnarray*}
m(D_{sJ}(2860)^+) = (2856.6 \pm 1.5 \pm 5.0) \ \mevcc \\
\Gamma(D_{sJ}(2860)^+)=(47 \pm 7 \pm 10) \ \mevcc.
\end{eqnarray*}

\begin{eqnarray*}
m(X(2690)^+)= (2688 \pm 4  \pm 3) \  \mevcc \\
\Gamma(X(2690)^+)=(112 \pm 7 \pm 36) \ \mevcc.
\end{eqnarray*}

In summary, in 240~${\rm fb}^{-1}$ of data collected by the
\babar\  experiment,  a new $D_s^+$  state is observed in the
inclusive $D K$ mass distribution near 2.86~\gevcc in three
independent channels. The decay to two pseudoscalar mesons implies
a natural spin-parity for this state: $J^P=0^+, 1^-,\dots$. It has
been suggested that this new state could be a radial excitation of
$D^*_{sJ}(2317)$~\cite{beveren} although other possibilities
cannot be ruled out. In the same mass distributions  a broad
enhancement around 2.69 \gevcc is also observed, it is not
possible to associate it to any known reflection or background.
This analysis is published~\cite{dsj}.
 Another \babar\
analysis\cite{vp}, has searched for resonances in \B \to
\D$^{(*)}$\D$^{(*)}$\K decays in 22 decay modes using 347~${\rm
fb}^{-1}$ data sample recorded at the $\Upsilon(4S)$ resonance.
The \D\K and \D$^*$\K invariant mass distributions are built with
8 decay modes each. Both distributions show a resonant enhancement
around 2700 \mevcc. However, due to an unknown structure at low
mass in the \D\K invariant mass distribution and to the possible
 additional resonances in the signal region in the \D$^*$\K
 invariant mass distribution, a full Dalitz analysis in necessary
 and is ongoing in order to extract the $D_{sJ}(2700)^+$
 parameters.

\section{Conclusion}

Although the nature of the newly discovered charm resonances is
not yet fully understood, the resonances are interpreted as
molecular or hybrid states in most theoretical papers. It will be
interesting to see if these interpretations are confirmed by
future measurements and analyses.

\bigskip 

\begin{thebibliography}{99}
\bibitem{footnote}Charge-conjugate reactions are implied throughout.
\end{thebibliography}

\end{document}